\documentclass[english,prb,aps,amsmath,amssymb,reprint,superscriptaddress,showpacs]{revtex4-1}
\usepackage[T1]{fontenc}
\usepackage[latin9]{inputenc}
\usepackage{color}
\usepackage{babel}
\usepackage{array}
\usepackage{units}
\usepackage{textcomp}
\usepackage{multirow}
\usepackage{amsmath}
\usepackage{amssymb}
\usepackage{graphicx}
\usepackage[unicode=true,
 bookmarks=true,bookmarksnumbered=false,bookmarksopen=false,
 breaklinks=true,pdfborder={0 0 0},pdfborderstyle={},backref=false,colorlinks=true,pdfpagemode=FullScreen]
 {hyperref}
\hypersetup{pdftitle={Atomic structure of the single step and dynamics of Sn adatoms on the Si(111)-sqrt(3)xsqrt(3)-Sn surface},
 pdfauthor={R. Zhachuk, D. Rogilo, A. Petrov, D. Sheglov, A. Latyshev, S. Colonna and F. Ronci},
 pdfsubject={68.35.Md, 68.43.Fg, 68.35.Ja},
 pdfkeywords={Silicon, Tin, Surface step, Surface reconstruction, Atom dynamics},
 allcolors=blue}
\usepackage{breakurl}

\makeatletter

\providecommand{\tabularnewline}{\\}

\makeatother

\begin{document}

\title{Atomic structure of the single step and dynamics of Sn adatoms on
the Si$(111)-\sqrt{3}\times\sqrt{3}-$Sn surface }

\author{R. A. Zhachuk}
\email{zhachuk@gmail.com}

\affiliation{Institute of Semiconductor Physics, pr. Lavrentyeva 13, Novosibirsk
630090, Russia}

\author{D. I. Rogilo}

\affiliation{Institute of Semiconductor Physics, pr. Lavrentyeva 13, Novosibirsk
630090, Russia}

\author{A. S. Petrov}

\affiliation{Institute of Semiconductor Physics, pr. Lavrentyeva 13, Novosibirsk
630090, Russia}

\author{D. V. Sheglov}

\affiliation{Institute of Semiconductor Physics, pr. Lavrentyeva 13, Novosibirsk
630090, Russia}

\author{A. V. Latyshev}

\affiliation{Institute of Semiconductor Physics, pr. Lavrentyeva 13, Novosibirsk
630090, Russia}

\author{S. Colonna}

\affiliation{CNR\textendash Istituto di Struttura della Materia, via del Fosso
del Cavaliere 100, I-00133 Roma, Italy}

\author{F. Ronci}

\affiliation{CNR\textendash Istituto di Struttura della Materia, via del Fosso
del Cavaliere 100, I-00133 Roma, Italy}

\date{\today}
\begin{abstract}
The atomic structure of well-ordered single steps on the Si$(111)-\sqrt{3}\times\sqrt{3}-$Sn
surface and the dynamics of Sn adatoms in the vicinity of these steps
were studied. The work was performed using low temperature scanning
tunneling microscopy (LT-STM) and\emph{ ab initio} calculations based
on the density functional theory. The STM tip was used to record the
tunneling current \emph{versus} time on top of oscillating adatoms,
keeping the feedback loop turned off. The dynamics of adatoms, detected
as the telegraph noise present in the tunneling current, was registered
near the steps at 80 K. The atomic structure model of the single steps
consisting of Sn atomic chains along the steps was developed. This
structure leads to the formation of potential double-wells near the
steps acting as traps for Sn atoms and explains the fluctuating current
recorded in these areas.
\end{abstract}


\maketitle

\section{INTRODUCTION }

Among other chemical elements used in silicon-based heterostructures
for micro- and photoelectronic applications, tin attracts particular
interest since it belongs to the same group as silicon and germanium.
Although the Si$(100)$ surface is a primary substrate for the fabrication
of functional heterostructures based on GeSn solid solutions \cite{wir16},
Si$(111)$ substrates provide a higher hole mobility for GeSn MOSFETs
\cite{mae15}. However, the epitaxial growth of high-quality GeSn
alloys on the Si surface is still a challenge because of the Sn surface
segregation and low solubility in the Ge matrix \cite{wir16,oeh13}.
On the Si$(111)$ surface, GeSn structures without the Sn segregation
are observed after the low-temperature growth of 5 nm amorphous GeSn
layer \cite{mae15} or after the initial Sn film growth followed by
the Ge deposition and substrate annealing \cite{yu16}. In the second
case, the Sn film starting from an even submonolayer thickness acts
as a surface active agent affecting the following epitaxial growth
on the Si surface \cite{dol08,lin95,iwa92,iwa92-2}. In particular,
the surfactant Sn layer enhances the surface diffusion and suppresses
the 2D island nucleation during the Ge/Si \cite{dol08} and Si/Si
\cite{iwa92,iwa92-2} epitaxy on the Si$(111)$ surface and prevents
the Ge segregation in near-surface layers of the Si substrate \cite{lin95}.
This modification of the atomic processes on the growing Si$(111)$
surface is related to the formation of Sn-induced surface reconstructions,
different from the intrinsic $7\times7$ dimer-adatom-stacking fault
structure \cite{tak85}, which change the nature of the interaction
between deposited atoms and the substrate. The presence of preliminarily
formed Sn-induced reconstructions on the Si$(111)$ surface also changes
the orientations of the epitaxial Sn layers growing at room temperatures
\cite{ryu02}. The investigation of the Sn/Si$(111)$ surface atomic
structure forming at the initial stages of Sn film growth is a way
to understand how to avoid the Sn segregation at the following stages
of GeSn structure fabrication.

Depending on the substrate temperature and deposited coverage, the
Sn adsorption on the Si(111)-7\texttimes 7 surface induces the formation
of the following ordered surface phases \cite{char01,nog89,tor94}:
$\alpha-\sqrt{3}\times\sqrt{3}$ ($\sqrt{3}\times\sqrt{3}$ shortly),
$\gamma-\sqrt{3}\times\sqrt{3}$ (mosaic phase) and $2\sqrt{3}\times2\sqrt{3}$.
The structure, atomic and electronic properties of these reconstructions
on the Si$(111)$ surface were well studied by the scanning tunneling
microscopy (STM) technique \cite{nog89,tor94,tor91,eri10,lev96,ich03}
and by\emph{ ab initio} calculations \cite{ich03,res04}. To reveal
the properties of a perfect reconstructed surface, these investigations
were carried out mostly on singular regions far from surface defects
and atomic steps always existing on real substrates. However, the
kinetics of epitaxial growth and structural transitions on the Si$(111)$
surface is directly determined by the features of atomic processes
near the steps \cite{lat92,kat00} depending on the step edge orientation
\cite{rom07,voi01,tey14,rog13}. One interesting consequence of this
is the reduced symmetry island growth on the higher symmetry substrate
demonstrated in case of the Si and Ge epitaxy on the Si$(111)-\sqrt{3}\times\sqrt{3}-$Bi
surface \cite{rom07}. In the present work, we used STM and\emph{
ab initio} calculations based on the density functional theory to
reveal the atomic structure of the steps on the Si$(111)-\sqrt{3}\times\sqrt{3}-$Sn
surface and to study the dynamics of adsorbed atoms in the vicinity
of such steps.

The paper is organized as follows: after reviewing the experimental
and calculation procedures, we report on the atomic structure model
of the single step on the Si$(111)-\sqrt{3}\times\sqrt{3}-$Sn surface,
give corresponding experimental and calculated STM images and show
the localization of empty and filled electronic states near the steps.
In Sec. 4.2 we first review the previous results related to the Sn
adatom dynamics on the flat $(111)-\sqrt{3}\times\sqrt{3}-$Sn surface
regions of Si and Ge. Then we report the STM data related to the adatom
dynamics on the stepped Si$(111)-\sqrt{3}\times\sqrt{3}-$Sn surface,
calculate the potential energy surfaces (PES) for Sn and Si adatoms,
compare the experimental and calculated data, and explain the fluctuating
current detected near the steps.

\section{METHODS}

\subsection{Experimental procedure}

The experiments were carried out using an ultra-high vacuum system
(base pressure $1\times10^{-10}\,\mathrm{\text{mbar}}$) equipped
with a low temperature scanning tunneling microscope (LT-STM, Omicron)
and a low energy electron diffraction (LEED). Silicon substrates were
cut from Si$(111)$ n-type wafers with resistivity $4\,\mathrm{m\varOmega\cdot cm}$.
The sample preparation started from clean Si$(111)$ surfaces. The
Si$(111)-7\times7$ clean surface was prepared by the annealing at
$900\,\mathrm{\text{\textdegree}C}$ and flashing the sample at $1250\,\mathrm{\text{\textdegree}C}$
for about $1\,\mathrm{min}$. The samples were resistively heated
by a direct current. The $7\times7$ surface reconstruction was confirmed
by LEED and STM before the Sn evaporation. Sn was evaporated from
a Knudsen effusion cell. The effusion cell was thoroughly outgassed
before its use in order to maintain the pressure in the $10^{-10}\,\mathrm{\text{mbar}}$
range during the metal deposition. The evaporation rate was measured
by using the quartz crystal thickness monitor. A nominal $\nicefrac{1}{3}$
monolayer Sn deposition was performed at room temperature, followed
by the sample annealing at $650\,\mathrm{\text{\textdegree}C}$. After
that the formation of the $\alpha-\sqrt{3}\times\sqrt{3}-$Sn reconstruction
was confirmed by LEED and STM. Electrochemically etched tungsten STM
tips were used after a cleaning procedure by electron bombardment.
The reported STM images were recorded at 80~K in the constant-current
mode. The STM scanner was calibrated by using the Si$(111)-7\times7$
surface as a reference.

Tunneling current vs time traces (\textquoteleft \textquoteleft current
traces\textquoteright \textquoteright{} hereafter) were acquired concomitantly
to the acquisition of constant current STM images by interrupting
the tip scan every five points and five lines over the chosen area.
At every single grid point, the STM feedback loop was switched off
and the tunneling current was recorded with a sampling rate of $33\,\mathrm{kHz}$.
In this way, we could observe steps in the current trace if an adatom
underneath the STM tip is unstable, i.e. moves horizontally or vertically.
Furthermore, it is possible to exactly locate the position of moving
adatoms on the STM images. To show where such traces were recorded
on the sample surface, we calculated the standard deviation ($\sigma$)
of all the current traces and reported these values on the \emph{z}
axis of a grid. The resulting images (hereinafter called \textquoteleft \textquoteleft $\sigma$
maps\textquoteright \textquoteright ) show a brighter color for higher
$\sigma$ values, evidencing where the adatom movement occurs. It
is important to emphasize that, with such an acquisition method, the
uncertainties of the actual tip location during the current trace
acquisition are greatly reduced because the STM image and the current
traces are acquired concomitantly. The maximum limit of the detectable
frequency in the current traces in our STM system is estimated to
be $\approx5\,\mathrm{kHz}$, while the lower limit is about $100\,\mathrm{Hz}$.
The method was successfully applied to the study of Sn atom dynamics
on flat Si$(111)-\sqrt{3}\times\sqrt{3}-$Sn and Ge$(111)-\sqrt{3}\times\sqrt{3}-$Sn
surfaces at low temperatures \cite{ron05,ron07} and was described
in detail in Ref.~\onlinecite{ron10}.

\subsection{Computational details}

The calculations were carried out using the pseudopotential \cite{tro91}
density functional theory \textsc{siesta} code \cite{sol02} within
generalized gradient approximation (GGA) to the exchange and correlation
interactions between electrons \cite{per96}. We did not consider
the corrections to GGA, which could help to better describe the electron
correlation effects in the Si$(111)-\sqrt{3}\times\sqrt{3}-$Sn system.
Such corrections, although having some influence on the calculated
electronic structure \cite{flo01,pro07,mod07,sch10}, are expected
to have a small impact on the calculated total energies and forces,
and the resulting atomic structure, studied in this work. The valence
states were expressed as linear combinations of the Sankey-Niklewski-type
numerical atomic orbitals \cite{sol02}. In the present calculations,
the polarized double-$\zeta$ functions were assigned for all atom
species. This means two sets of \emph{s}- and \emph{p}-orbitals plus
one set of \emph{d}-orbitals on silicon and tin atoms, and two sets
of \emph{s}-orbitals plus a set of \emph{p}-orbitals on hydrogen atoms.
The electron density and potential terms were calculated on a real
space grid with the spacing equivalent to a plane-wave cut-off of
200 Ry.

In our calculations we used (111)-slabs with single steps in the $\left[11\overline{2}\right]$
step-down direction ($\left[11\overline{2}\right]$-steps in short).
The dangling bonds at the slabs bottom side were saturated by H atoms,
while the top $(111)$-terraces were constructed according to the
$\sqrt{3}\times\sqrt{3}-$Sn atomic model consisting of Sn-adatoms
on $\mathrm{T_{4}}$-sites \cite{nog89}. The slab thickness was about
$14$~Å (four Si-bilayers), and the $26$~Å thick vacuum layer was
used. The positions of all slab atoms (except for the Si atoms in
the bottom bilayer and all H atoms) were fully optimized until the
atomic forces became less than $0.01$~eV/Å. Two different calculation
cells were used in this work. For the calculation of relative formation
energies and PES maps, we used slabs with $11.6\times18.2\,\mathrm{\mathring{A}^{2}}$
lateral dimensions and a grid of $4\times3\times1$ $\mathbf{k}$-points
in the Brillouin zone \cite{mon76}. Larger slabs with $11.6\times58.4\,\mathrm{\mathring{A}^{2}}$
lateral dimensions and a grid of $4\times1\times1$ $\mathbf{k}$-points
were used for the calculation of STM images. Si and Sn chemical potentials
were calculated using bulk unit cells and $10\times10\times10$ $\mathbf{k}$-point
grids. The Sn chemical potential was calculated using the bulk $\alpha$-phase
of Sn having the diamond structure and known as gray tin. The calculated
Si and Sn bulk-lattice constants are $a_{Si}=5.50\,\mathrm{\mathring{A}}$
and $a_{Sn}=6.73\,\mathrm{\mathring{A}}$, which are close to experimental
values $a_{Si}=5.43\,\mathrm{\mathring{A}}$ and $a_{Sn}=6.46\,\mathrm{\mathring{A}}$,
respectively. The constant-current STM images were produced based
on the Tersoff-Hamann approximation \cite{ter85} using the eigenvalues
and eigenfunctions of the Kohn-Sham equation \cite{koh65} for a relaxed
atomic structure. Experimental and calculated STM images were processed
using the \textsc{WSxM} software \cite{hor07}. 

In this work, we compared the relative formation energies of the stepped
Si$(111)-\sqrt{3}\times\sqrt{3}-$Sn surface according to different
atomic configurations of the step edge. The Si$(111)-\sqrt{3}\times\sqrt{3}-$Sn
surface having an unreconstructed step edge (described in the text)
after the structure relaxation was treated as a reference surface.
The relative surface formation energies (per unit area) were calculated
according to the procedure described in Ref.~\onlinecite{bat08}:

\begin{equation}
\Delta\gamma=\left(E_{model}-E_{unrec}-\mu_{Si}\Delta N_{Si}-\mu_{Sn}\Delta N_{Sn}\right)/S,\label{eq:1}
\end{equation}
where $S$ is the unit cell area of the slab in the \emph{xy} plane,
$E_{unrec}$ refers to the total energy of the reference slab containing
the unreconstructed step after the structure relaxation, $E_{model}$
is the total energy of the slab according to the trial step model,
$\Delta N_{Si}$ and $\Delta N_{Sn}$ account for the number of Si
and Sn atoms in the trial model in excess of those in the reference
model. 

$\mu_{Si}$ and $\mu_{Sn}$ are the chemical potentials of Si and
Sn, which are the energy per atom in the reservoirs with which the
surface is assumed to be in equilibrium. Since the surface is in equilibrium
with the bulk Si substrate, $\mu_{Si}$ is the energy per atom in
bulk Si. The adsorbate chemical potential $\mu_{Sn}$ was treated
as a variable, since its exact value is unknown. It corresponds to
a real physical variable that can be externally tuned \cite{bat08}.
Under thermodynamic equilibrium conditions the surface having the
lowest formation energy will be realized. The change of $\mu_{Sn}$
leads to a change of surface formation energy, which may be followed
by structural phase transitions. The chemical potential in bulk $\alpha$-phase
Sn ($\mu_{Sn-bulk}$) may be considered as the upper bound for the
Sn chemical potential. If $\mu_{Sn}>\mu_{Sn-bulk}$, then bulk $\alpha$-phase
Sn would be energetically more favorable than any adsorbed phase,
which would lead to the precipitation of bulk Sn on the Si surface.
In the opposite case of extreme low $\mu_{Sn}$ values, the clean
Si surface with no Sn adatoms becomes the most stable. Thus, the low/high
$\mu_{Sn}$ values should favor atomic structures with a low/high
amount of Sn atoms on the Si surface, respectively. In this work we
are interested in the intermediate $\mu_{Sn}$ values, when the Sn-Si
compound surface phases are stable.

The PES maps describing the energetics of adatoms on the stepped Si$(111)-\sqrt{3}\times\sqrt{3}-$Sn
surface were produced for the positions of the adsorbed probe Sn (Si)
atom within a symmetry-irreducible half of the calculation cell. For
each \emph{xy}-coordinate, the adsorbed atom was initially placed
approximately $3$~Å above the surface and its \emph{z}-coordinate
was allowed to relax, while the coordinates in the \emph{xy}-plane
were kept fixed. The positions of all remaining atoms, but the bottom
Si-H units, were fully optimized until atomic forces became less than
$0.01$~eV/Å. The total of 444 points forming a grid with about $0.5$~Å
spacing were calculated in the calculation cell half. The PES data
for the whole calculation cell were recovered by applying the mirror
symmetry transformation to the calculated PES data. The exact energies
of local energy minima on PES were calculated with a free-moving probe
atom placed near the local energy minima.

\section{RESULTS AND DISCUSSION}

\subsection{Atomic structures of the single step on the Si$(111)-\sqrt{3}\times\sqrt{3}-$Sn
surface}

\begin{figure}
\includegraphics[clip,width=8.5cm]{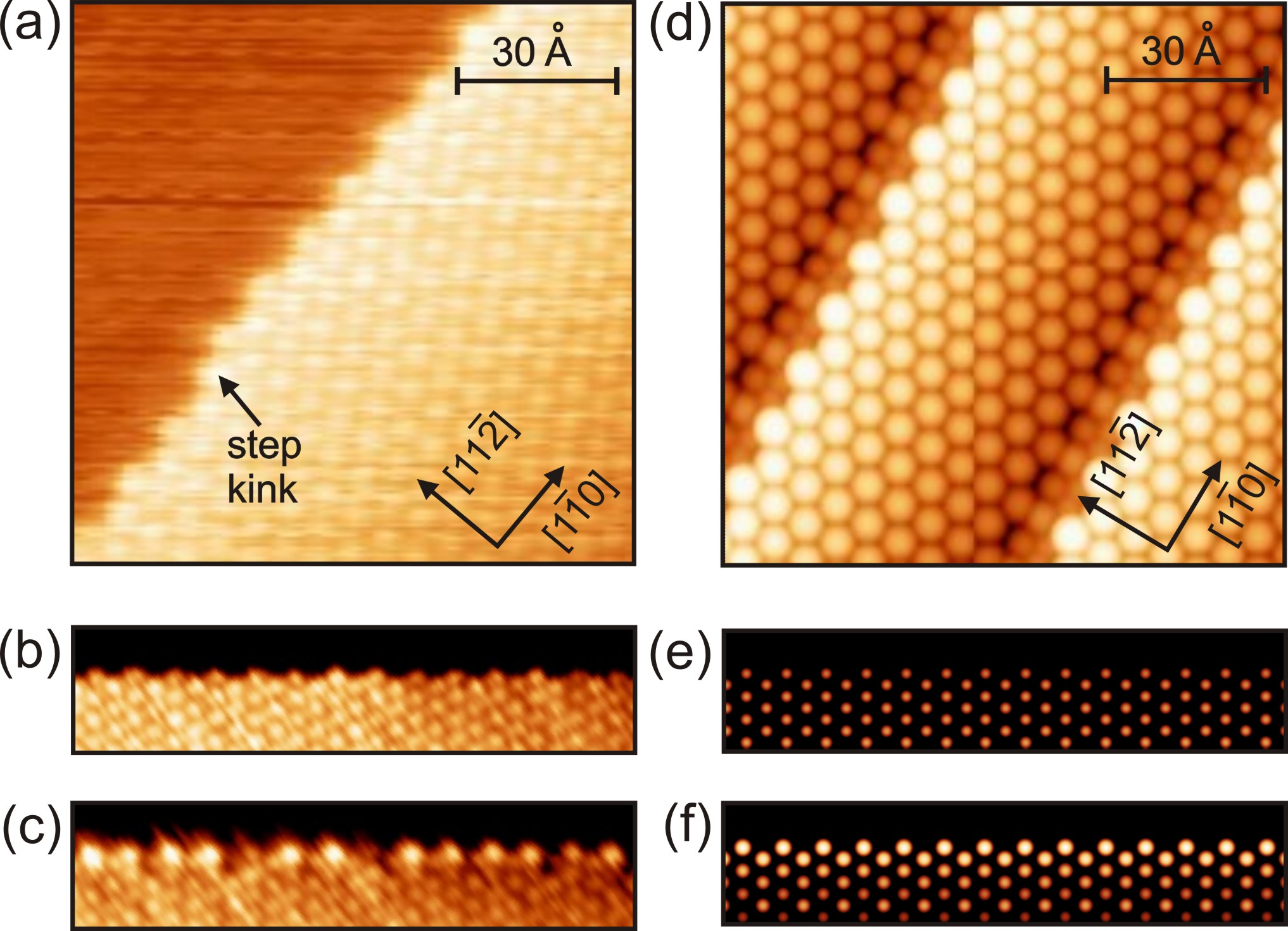}

\caption{\label{fig1} Experimental (a), (b), (c) and calculated (d), (e),
(f) constant-current STM images of single steps on the Si$(111)-\sqrt{3}\times\sqrt{3}-$Sn
surface. (a), (b), (d), (e) $U=+1.0\,\mathrm{V}$, (c), (f) $U=-1.0\,\mathrm{V}$.
The calculated STM images are based on the \textquotedblleft B\textquotedblright{}
atomic model of the single step on the Si$(111)-\sqrt{3}\times\sqrt{3}-$Sn
surface (see the text).}
\end{figure}

The starting point of our investigation of the step atomic structure
is the information provided by STM images. In Fig.~\ref{fig1}(a)
is a typical STM-image of the atomic step found on the Si$(111)-\sqrt{3}\times\sqrt{3}-$Sn
surface. Each bright spot in the STM image represents a Sn adatom.
In Figs.~\ref{fig1}(b) and \ref{fig1}(c) are the high-resolution
STM-images of the step edge obtained at $U=+1.0\,\mathrm{V}$ and
$U=-1.0\,\mathrm{V}$, respectively. It is seen in Fig.~\ref{fig1}(c)
that the Sn adatoms at the upper terrace edge look brighter than those
far from it at $U=-1.0\,\mathrm{V}$. Since this effect was observed
only at a negative bias, then it is unlikely that it is a real topography
feature. The effect is most probably due to a charge transfer and
associated higher density of filled local electron states (electrons)
at the Sn atoms located at the upper terrace edge.

\begin{figure}
\includegraphics[clip,width=8cm]{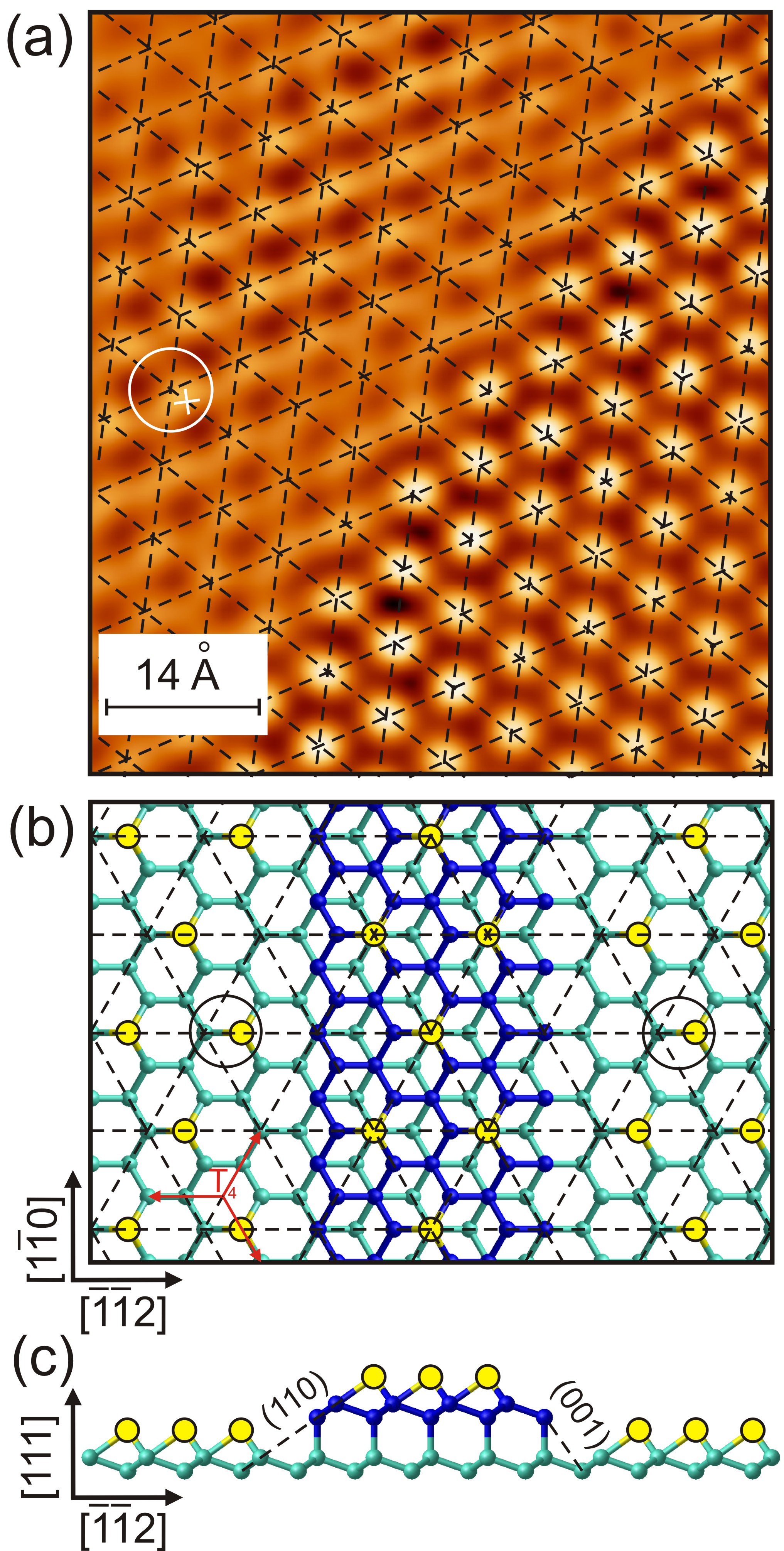}

\caption{\label{fig2} (a) Experimental STM image of the step on the Si$(111)-\sqrt{3}\times\sqrt{3}-$Sn
surface after FFT band pass filtering, $U=-1.0\,\mathrm{V}$. The
image was carefully corrected for the minimal distortion using the
hexagonal lattice of the upper terrace (lower right part of the image)
as a reference. The white cross marks the position of bright spot
on the lower terrace. (b), (c) Atomic models of unreconstructed steps
on the Si$(111)-\sqrt{3}\times\sqrt{3}-$Sn surface with $\left[11\overline{2}\right]$
and $\left[\overline{1}\overline{1}2\right]$ steps (left and right
step, respectively). Cyan and blue circles are Si atoms of the lower
and upper $(111)$ terraces, respectively, yellow circles are Sn adatoms.
The hexagonal lattices (dashed lines) in (a) and (b) are drawn across
the positions of Sn adatoms on the upper terrace, while large circles
highlight the deviations of the lattice nodes from the nearest Sn
positions.}
\end{figure}

In Fig.~\ref{fig2}(a) is the experimental STM image of the single
step on the Si$(111)-\sqrt{3}\times\sqrt{3}-$Sn surface acquired
at $U=-1.0\,\mathrm{V}$ after a linear distortion correction and
an FFT band pass filtering. The key for understanding the step edge
directions is in the mutual alignment of the $\sqrt{3}\times\sqrt{3}-$Sn
lattices on the neighboring $(111)$ terraces separated by the single
step. Accordingly, the hexagonal lattice was drawn across the Sn adatom
positions on the upper terrace (lower right part of Fig.~\ref{fig2}(a).
It is clear that the Sn adatoms on the lower terrace (upper left part
of Fig.~\ref{fig2}(a)) are located on the same lines drawn perpendicular
to the step edge as the Sn adatoms on the upper terrace. Two (the
only possible) models of unreconstructed steps on the Si$(111)-\sqrt{3}\times\sqrt{3}-$Sn
surface compatible with the above observation are shown in Figs.~\ref{fig2}(b)
and \ref{fig2}(c). Thus, the direction along the step edge in Figs.~\ref{fig1}(a)
and \ref{fig2}(a) is of $\left[1\overline{1}0\right]$-type, as it
follows from the models shown in Figs.~\ref{fig2}(b) and \ref{fig2}(c).

A more difficult question is the step-down direction in the experimental
STM images shown in Figs.~\ref{fig1}(a) and \ref{fig2}(a). This
direction can be either $\left[\bar{1}\overline{1}2\right]$, resulting
in a step in which the bonds configuration is the same as on unreconstructed
$(001)$ surface (right parts of Figs.~\ref{fig2}(b) and \ref{fig2}(c))
or $\left[11\overline{2}\right]$, in which the bonds configuration
is as on the $(110)$ surface (left parts of Figs.~\ref{fig2}(b)
and \ref{fig2}(c)). It should be noted that these two directions
are physically nonequivalent for crystals with a diamond lattice, i.e.
they lead to a different step atomic structure with different properties.
To solve this problem, we note that the bright spots (Sn atoms) on
the lower terrace shown in Fig.~\ref{fig2}(a) are shifted from the
nearest nodes of the hexagonal lattice by about $\nicefrac{1}{3}$
lattice period ($\thickapprox2.2$~Å) towards the step edge. The
shift is clearly visible near the central part of the lower terrace.
The Sn atom positions on the lower terrace near the step edge should
be ignored since, in this area, the image may be altered by an undesired
tunneling through the side of the STM tip. The same shift direction
can be observed only for $\left[11\overline{2}\right]$ steps. Therefore,
the steps observed in our experimental STM images in Figs.~\ref{fig1}(a)
and \ref{fig2}(a) are $\left[11\overline{2}\right]$ steps.

The trial atomic models of the steps for \emph{ab initio} calculations
were constructed in accordance with the information provided by experimental
STM images. In this work, we compare the stability of different step
configurations by calculating the relative formation energies of regular
stepped surfaces according to various step models. The step formation
energies were not calculated explicitly. 

\begin{figure}
\includegraphics[clip,width=8.5cm]{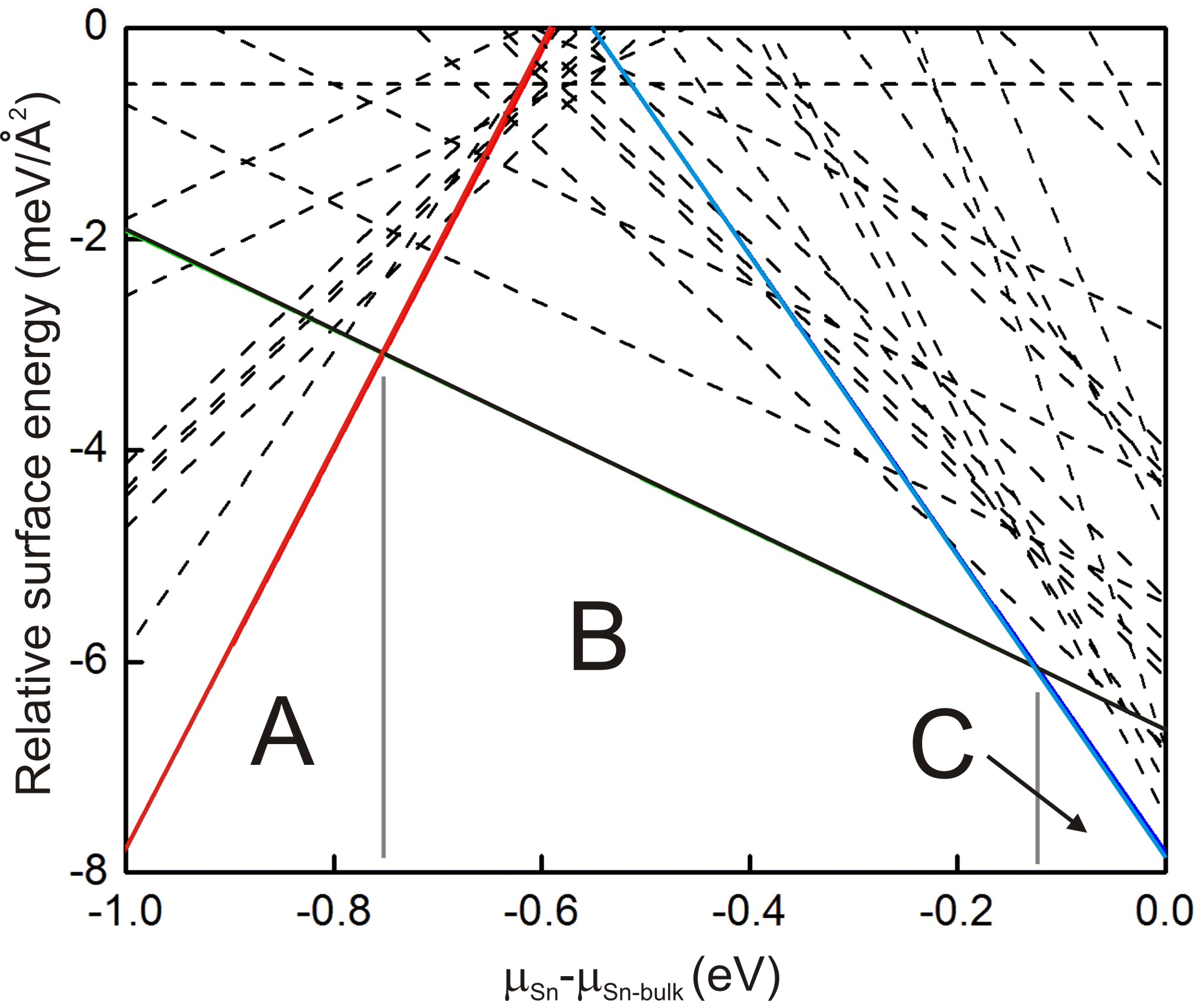}

\caption{\label{fig3} Surface-energy diagram of the stepped Si$(111)-\sqrt{3}\times\sqrt{3}-$Sn
surface. The energies are given relative to the formation energy of
relaxed surface with unreconstructed steps (see the text) for the
Sn adsorbate. The models marked as \textquotedblleft A\textquotedblright{}
(red solid line), \textquotedblleft B\textquotedblright{} (black solid
line), and \textquotedblleft C\textquotedblright{} (blue solid line)
correspond to the lowest-energy surface atomic configurations found
for $-1.0\,\mathrm{eV<\mu_{Sn}-\mu_{Sn-bulk}<0.0\,eV}$. Black dashed
lines are from other trial models. }
\end{figure}

We performed an extensive search for the lowest energy atomic configuration
of the stepped Si$(111)-\sqrt{3}\times\sqrt{3}-$Sn surface. 50 atomic
models were evaluated, and our results are shown in Fig.~\ref{fig3}.
Three solid lines - red, black, and blue - represent the surface energy
graphs for three lowest energy atomic configurations, A, B, and C,
depending on the $\mu_{Sn}$ value, while the dashed lines in Fig.~\ref{fig3}
are from other higher energy trial models.

\begin{figure}
\includegraphics[width=6.5cm]{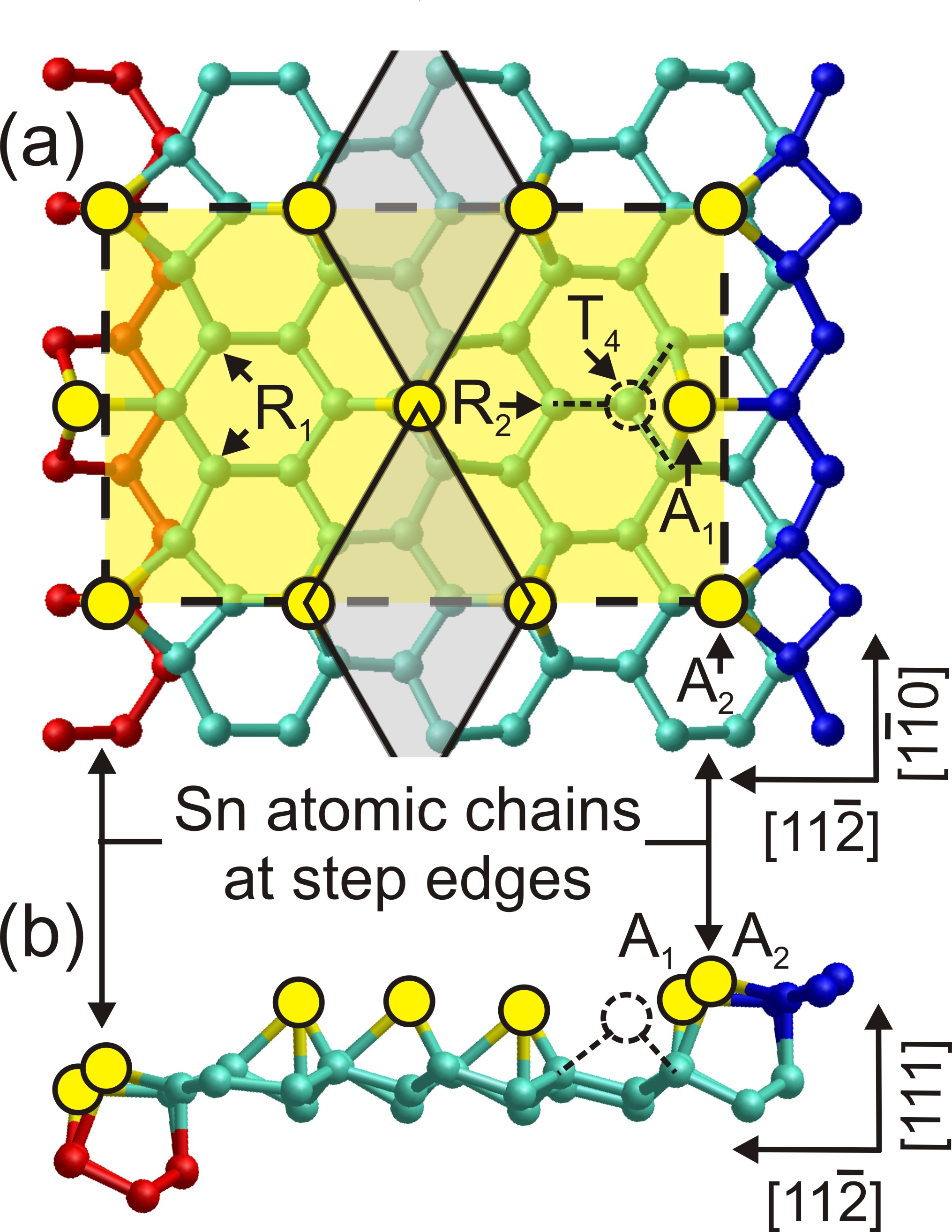}\caption{\label{fig4} Atomic model \textquotedblleft B\textquotedblright{}
of the step edge on the Si$(111)-\sqrt{3}\times\sqrt{3}-$Sn surface.
Red, cyan, and blue circles are Si atoms of the lower, middle and
upper $(111)$ terraces respectively, yellow circles are Sn adatoms.
Gray shaded rhombuses are the unit cells of the $\sqrt{3}\times\sqrt{3}-$Sn
reconstruction, the yellow shaded rectangle outlined by dashed line
is the calculation unit cell. $\mathrm{A_{1}}$ and $\mathrm{A_{2}}$
denote Sn adatoms at the step edges forming Sn atomic chains, $\mathrm{R_{1}}$
and $\mathrm{R_{2}}$ are the Si rest atoms. The site outlined by
a dashed circle and marked as $\mathrm{T_{4}}$ shows the place where
the $\mathrm{A_{1}}$ atom came from. (a) Top view. (b) Side view. }
\end{figure}

The atomic step model \textquotedblleft B\textquotedblright , which
is stable in a wide range of $\mu_{Sn}$ values (Fig.~\ref{fig3})
is shown in Fig.~\ref{fig4}. The difference between this model and
the model of unreconstructed step shown in Figs.~\ref{fig2}(b) and
\ref{fig2}(c) (left parts of these figures) is the presence of Sn
atomic chain consisting of $\mathrm{A_{1}}$ and $\mathrm{A_{2}}$
Sn adatoms decorating the step edge. The adatom $\mathrm{A_{1}}$
can be considered as the one taken from the $\sqrt{3}\times\sqrt{3}-$Sn
structure on the lower terrace by shifting from the neighboring $\mathrm{T_{4}}$
site to the step edge (see the dashed circle outlining the previous
position of this adatom). The adatom $\mathrm{A_{2}}$ is an additional
Sn atom, not existing in the unreconstructed step model. The model
\textquotedblleft B\textquotedblright{} contains Si atoms with unsaturated
dangling bonds (rest atoms): $\mathrm{R_{1}}$ on the upper terrace
and $\mathrm{R_{2}}$ on the lower terrace. The rest atoms show the
$sp^{2}$-like (planar) configuration of saturated bonds after the
structure relaxation. While $\mathrm{R_{1}}$ exists also in the unreconstructed
step model, the emergence of $\mathrm{R_{2}}$ is caused by the Sn
atom shift leading to the formation of $\mathrm{A_{1}}$ step edge
atom. All other Si bonds in the model, except for the bonds on $\mathrm{R_{1}}$
and $\mathrm{R_{2}}$ atoms, are saturated. It is difficult to resolve
$\mathrm{A_{1}}$,$\mathrm{A_{2}}$, $\mathrm{R_{1}}$ and $\mathrm{R_{2}}$
atoms experimentally since Sn adatoms, in the vicinity of the upper
terrace edge, strongly hinder their imaging. The possible reason of
why Sn atoms prefer $\mathrm{A_{1}}$ and $\mathrm{A_{2}}$ positions
at the step edges is the bigger size of Sn atoms, as compared to that
of Si. As a result, Sn atoms better fit the positions with large distances
to the closest neighbors, as it is the case for $\mathrm{A_{1}}$
and $\mathrm{A_{2}}$.

Model \textquotedblleft C\textquotedblright{} is stable at the excess
of Sn on Si surface, just before precipitation of the bulk Sn (Fig.~\ref{fig3}).
The only difference between \textquotedblleft B\textquotedblright{}
and \textquotedblleft C\textquotedblright{} models is that, in case
of the \textquotedblleft C\textquotedblright{} model, the Si rest
atom $\mathrm{R_{2}}$ is substituted with the Sn atom. Both models
result in very similar calculated STM images. Model \textquotedblleft A\textquotedblright{}
is stable at low $\mu_{Sn}$ values and contains no Sn atoms at all.
This clean Si surface model is basically the unreconstructed step
edge model shown in Figs.~\ref{fig2}(b) and \ref{fig2}(c) (left
parts) where all Sn atoms on $(111)$ terraces are substituted by
Si atoms. Thus, the terraces in model \textquotedblleft A\textquotedblright{}
are structurally similar to the mosaic $\sqrt{3}\times\sqrt{3}-$Sn
$\gamma$-phase observed after the partial self-cleaning of Si$(111)$
terraces at elevated temperatures \cite{char01,tor94}. The models
\textquotedblleft A\textquotedblright{} and \textquotedblleft C\textquotedblright{}
determine the upper and lower stability boundaries of the \textquotedblleft B\textquotedblright{}
model, which is most likely to be relevant for the experimentally
observed step structure.

In Fig.~\ref{fig1}(d) is a large-scale calculated STM image of the
Si$(111)-\sqrt{3}\times\sqrt{3}-$Sn surface based on the atomic step
model \textquotedblleft B\textquotedblright{} shown in Fig.~\ref{fig4}.
In Figs.~\ref{fig1}(e) and \ref{fig1}(f) are the STM images of
the step edges calculated for $U=+1.0\,\mathrm{V}$ and $U=-1.0\,\mathrm{V}$,
respectively. The latter two images exhibit the same features as the
experimental STM images shown in Figs.~\ref{fig1}(b) and \ref{fig1}(c):
namely, the spots associated with Sn adatoms at the very edge of the
upper terrace are brighter than other spots on the filled states STM
images ($U<0$ ). This effect was not observed at $U=+1.0\,\mathrm{V}$
in agreement with the experimental data (Figs.~\ref{fig1}(b) and
\ref{fig1}(e)). Thus, the experimental and calculated STM images
demonstrate a good match, which is a prerequisite for a correct atomic
model. It should be noted that few other trial models exhibited the
same effect of bright terrace edges as the model shown in Fig.~\ref{fig4},
but these atomic configurations had significantly higher formation
energies. 

\begin{figure}
\includegraphics[width=7.5cm]{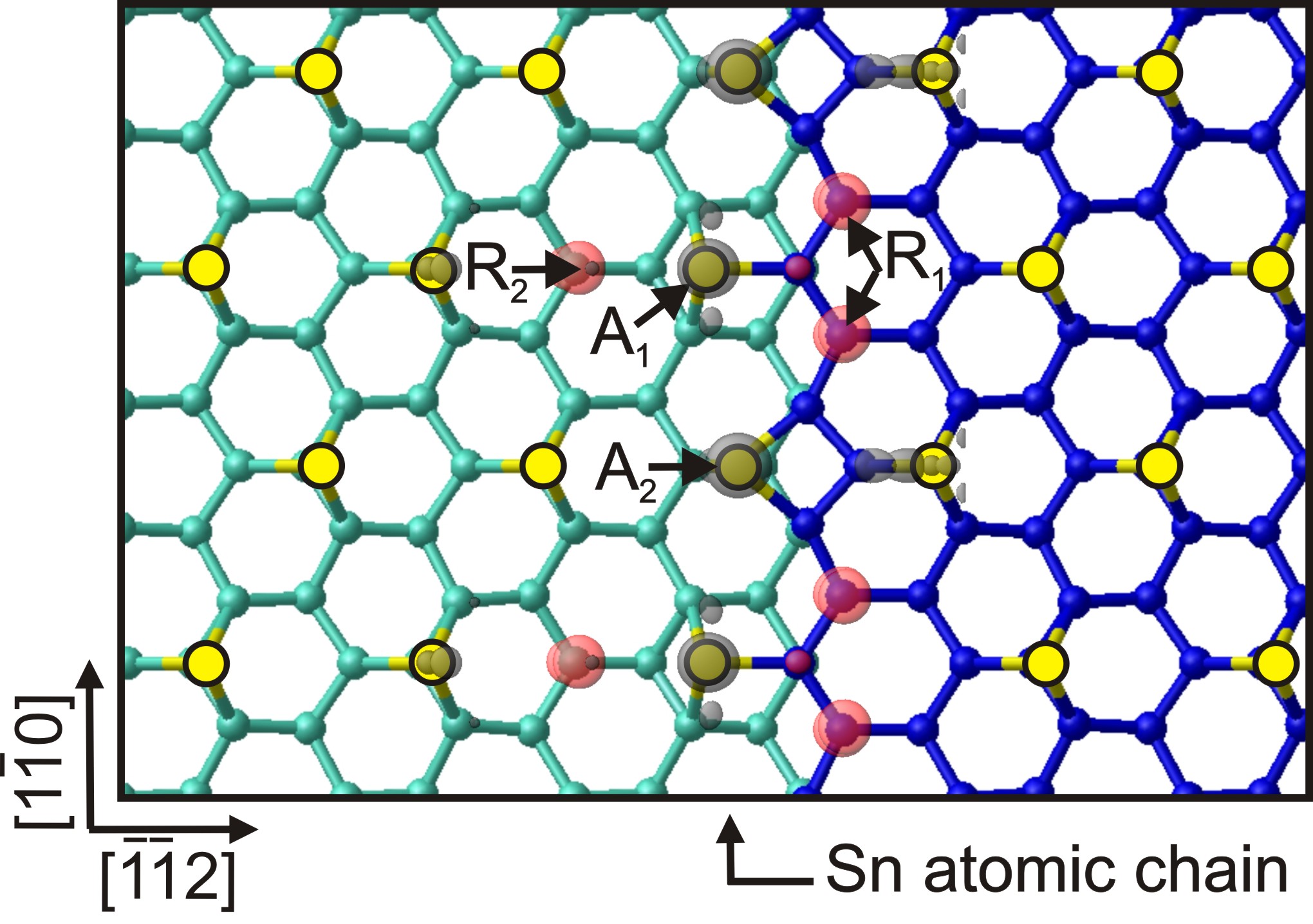}

\caption{\label{fig5}LDOS isosurfaces integrated in a $1.0\,\mathrm{eV}$
energy window below (half-transparent gray areas) and above (half-transparent
red areas) the calculated Fermi level. The atomic model \textquotedblleft B\textquotedblright{}
of the Si$(111)-\sqrt{3}\times\sqrt{3}-$Sn stepped surface is overlaid.
Cyan and blue circles are Si atoms of the lower and upper $(111)$
terraces, respectively, yellow circles are Sn adatoms. $\mathrm{A_{1}}$
and $\mathrm{A_{2}}$ denote Sn adatoms at the step edge forming the
Sn atomic chain, $\mathrm{R_{1}}$ and $\mathrm{R_{2}}$ are the Si
rest atoms on the upper and lower $(111)$ terraces, respectively. }
\end{figure}

The charge distribution near the step edge is the reason for the increased
intensity on Sn adatoms in the filled states experimental and calculated
STM images. This effect was investigated in more detail. The calculated
local density of states (LDOS) isosurfaces integrated in a 1.0~eV
energy window below and above the calculated Fermi level are shown
in Fig.~\ref{fig5}. The atomic model superimposed on this pattern
allows identifying the atoms acting as electron donors or acceptors.
It is clear that the empty states are mostly concentrated on the Si
rest atoms, while filled states are mostly located on $\mathrm{A_{1}}$
and $\mathrm{A_{2}}$ atoms of the Sn chain. Few filled states are
also found on the Sn atoms near the lower and upper terrace edges.

\subsection{PES and dynamics of Sn adatoms on the Si$(111)-\sqrt{3}\times\sqrt{3}-$Sn
surface with steps}

The dynamical fluctuations of Sn atoms on flat Sn/Si$(111)$ and Sn/Ge$(111)$
surfaces were studied in detail in Refs.~\onlinecite{ron05} and
\onlinecite{ron07} by recording the tunneling current as a function
of time. Vertical fluctuations of Sn atoms within the $\sqrt{3}\times\sqrt{3}$
surface reconstruction are related to the $\sqrt{3}\times\sqrt{3}\rightarrow3\times3$
phase transition during the temperature decrease. The process is well
described within the dynamical fluctuations model, at least in case
of the Sn/Ge$(111)$ system. According to this model, the static $\sqrt{3}\times\sqrt{3}$
structure is unstable and transforms into the $3\times3$ structure
with the Sn adatoms buckled up and down \cite{avi99}. The observation
of the $\sqrt{3}\times\sqrt{3}$ structure, according to this theory,
is due to the time averaged vertical fluctuations of Sn atoms, which
is $0.36$~Å between the up-and-down positions on Sn/Ge$(111)$ \cite{gor09}.
The fluctuations were observed at $80-220\,\mathrm{\text{K}}$ in
the Sn/Ge$(111)$ system and at $2.3-32\,\mathrm{\text{K}}$ in the
Sn/Si$(111)$ system. While the stable $3\times3$ structure was actually
observed in the Sn/Ge$(111)$ system at $80\,\mathrm{\text{K}}$,
the $\sqrt{3}\times\sqrt{3}$ reconstruction on Sn/Si$(111)$ persists
even at a very low temperature, down to $2.3\,\mathrm{\text{K}}$
\cite{ron10}. The reason for that is a very low energy barrier ($2.6\pm1.4\,\mathrm{meV}$
\cite{ron07}) associated with the phase transition on Sn/Si$(111)$,
which is much lower than that on Sn/Ge$(111)$ ($13\pm7\,\mathrm{meV}$
\cite{ron05}). This results in the notable quantum tunneling of Sn
adatoms between two stable positions on Sn/Si$(111)$ and that destroys
the $3\times3$ structural order \cite{ron07}.

Due to the very low energy barrier in the Sn/Si$(111)$ system, the
frequency of Sn dynamical fluctuations at $80\,\mathrm{\text{K}}$
is expected to be well above the maximum limit of detectable frequency
in our STM measurements (about $5\,\mathrm{\text{kHz}}$). It is possible
to estimate how high the energy barriers have to be in order to observe
the dynamical fluctuations in STM. To this end, we make use of the
frequency ($f$) of thermally activated dynamical fluctuations using
an Arrhenius relation $f=f_{0}exp\left(-E_{a}/kT\right)$, where $k$,
$T$, and $f_{0}$ stand for the Boltzmann constant, sample temperature
and attempt frequency, respectively, which can be approximated by
the Debye frequency of Si ($14\,\mathrm{THz}$). Thus, for the upper
limit of detectable frequency ($5\,\mathrm{kHz}$), we get $E_{a}=0.13\,\mathrm{eV}$
and for the lower limit ($100\,\mathrm{Hz}$) - $E_{a}=0.15\,\mathrm{eV}$. 

\begin{figure}
\includegraphics[width=6.5cm]{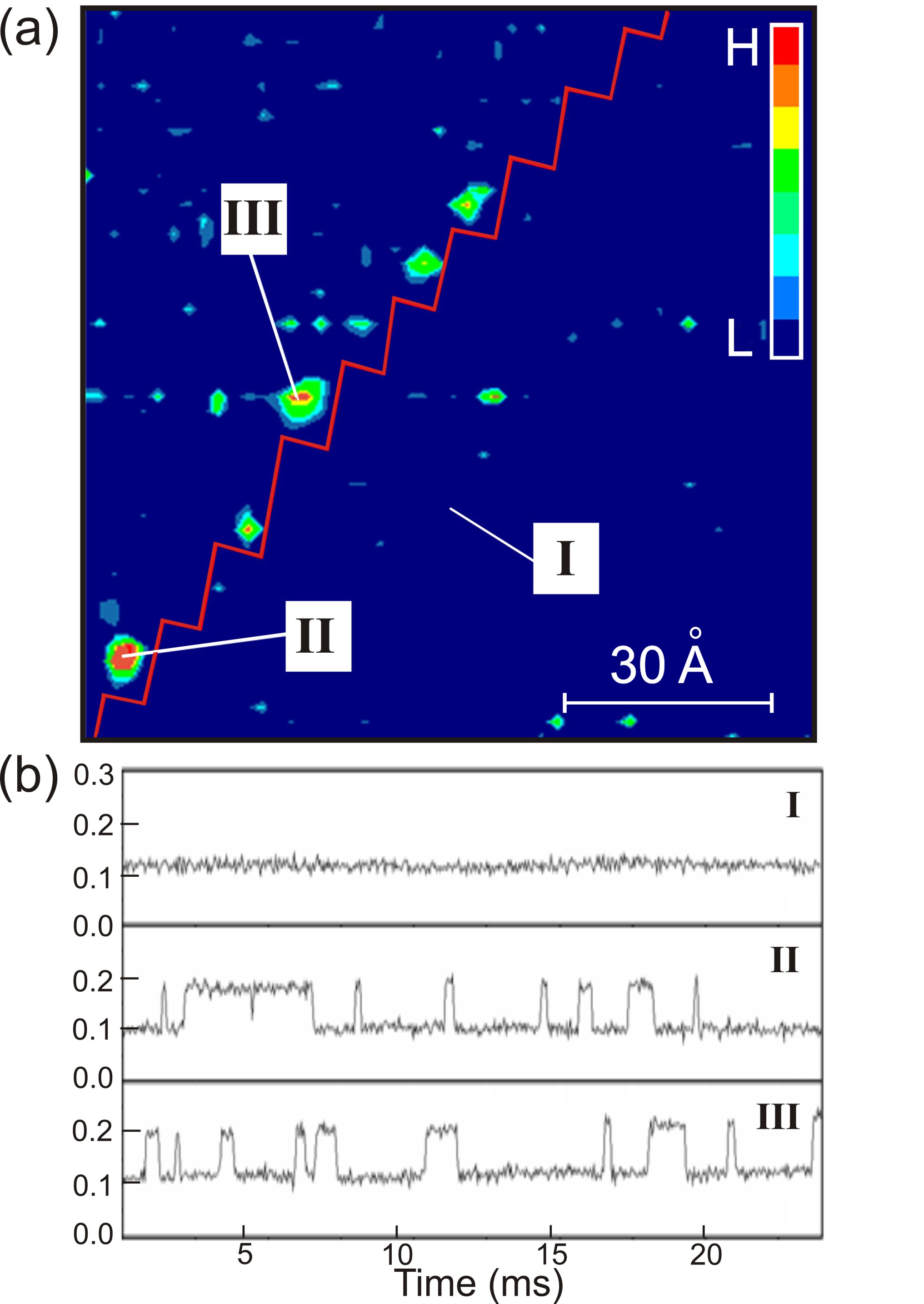}

\caption{\label{fig6}(a) $\sigma$ map related to the constant-current STM
image shown in Fig.~\ref{fig1}(a). The location of the step edge
in the STM image is indicated by a red line. (b) A collection of current
traces detected on the flagged surface spots.}
\end{figure}

Figure~\ref{fig6}(a) shows the $\sigma$ map acquired concomitantly
with the STM image shown in \ref{fig1}(a). All the current traces
recorded on the terrace show a flat profile ($\sigma\approx0$) consistent
with the expected high frequency of Sn dynamical fluctuations at 80~K
(see the current trace collected at point I, for example). However,
the $\sigma$ map clearly exhibits bright spots at the lower side
of step edge, that show the presence of current traces with a high
standard deviation value. As an example, two stepped current traces
collected at points II and III are reported in Fig.~\ref{fig6}(b).
The observation of stepped current traces near the step edge at $80\,\mathrm{\text{K}}$
indicates that the energy barrier for the fluctuating adatoms in that
area is much larger than that on the flat Sn/Si$(111)$ surface. It
is also worth noting that, in spite of a very regular step structure
visible in the STM image in Fig.~\ref{fig1}(a), only a few irregular
spots are visible in the $\sigma$ map in Fig.~\ref{fig6}(a). The
reason for that will be clear from the results presented below.

\begin{figure}
\includegraphics[width=7.5cm]{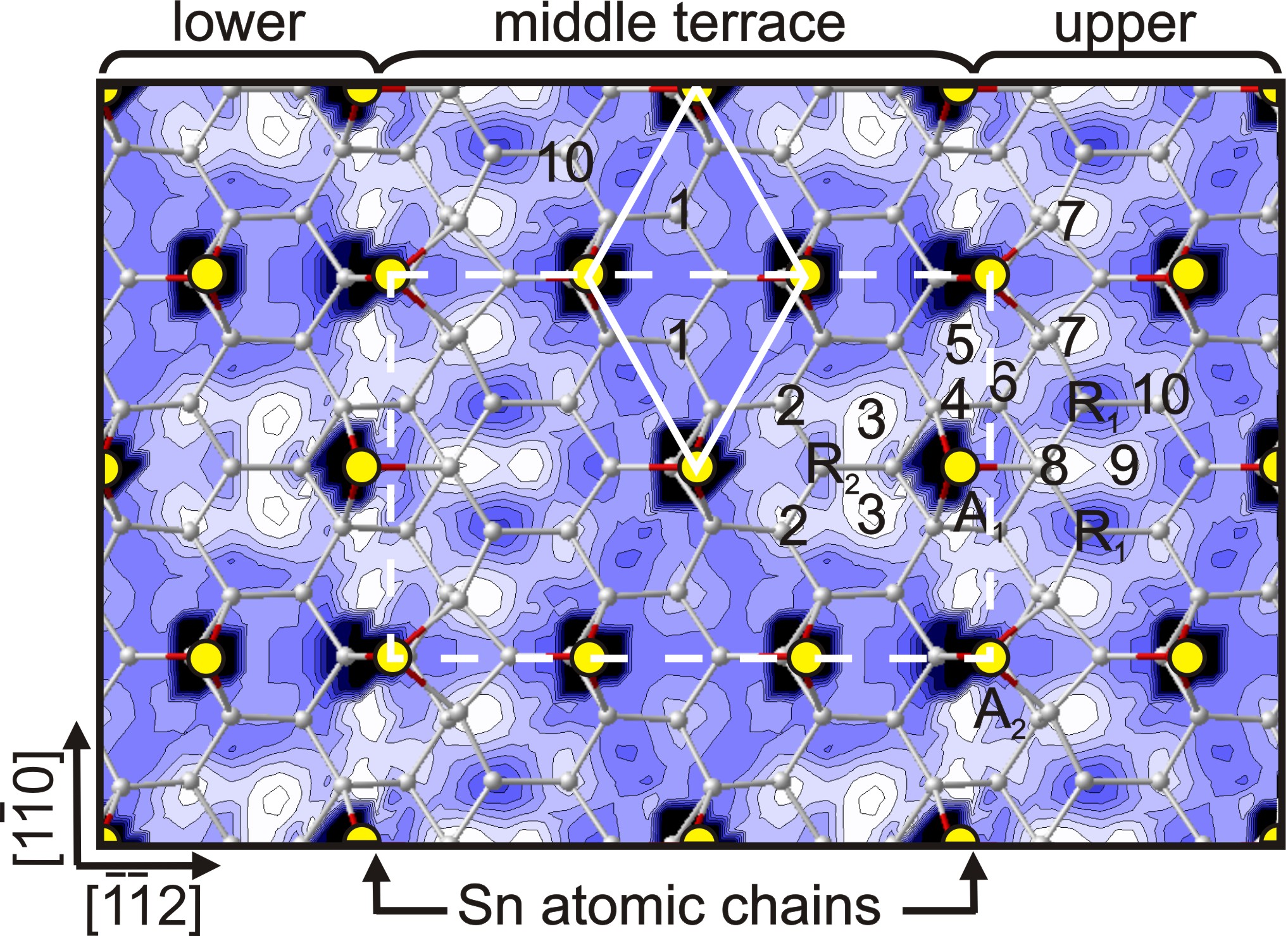}

\caption{\label{fig7}PES of an adsorbed Sn probe atom on a stepped Si$(111)-\sqrt{3}\times\sqrt{3}-$Sn
surface; the atomic model \textquotedblleft B\textquotedblright{}
is overlaid. The contour spacing is $0.2\,\mathrm{eV}$. Bright (dark)
regions indicate energy minima (maxima). Numbers indicate different
local minima on PES. White circles are Si atoms, yellow circles are
Sn atoms. $\mathrm{A_{1}}$ and $\mathrm{A_{2}}$ denote Sn atoms
at the step edges forming Sn atomic chains, $\mathrm{R_{1}}$ and
$\mathrm{R_{2}}$ are the Si rest-atoms on $(111)$ terraces. The
rectangle outlined by a white dashed line is the calculation unit
cell of the stepped surface, the rhombus outlined by a white solid
line is the $\sqrt{3}\times\sqrt{3}-$Sn unit cell on the $(111)$
terrace. }
\end{figure}

The calculated PES for a Sn probe atom on the stepped Si$(111)-\sqrt{3}\times\sqrt{3}-$Sn
surface with the atomic model \textquotedblleft B\textquotedblright{}
overlaid is shown in Fig.~\ref{fig7}. Bright regions correspond
to the energy minima, while dark regions indicate energy maxima. It
is clear that the Sn adatom positions on the surface are the highest
energy maxima on PES. This is because these atoms have only one dangling
bond and, therefore, these sites are not favorable for the adsorption
of additional probe Sn atom having four valence electrons. The positions
of rest atoms also show energy maxima, but their values are lower
than those found on Sn atoms. 

\begin{table}
\begin{ruledtabular}
\caption{\label{tab1}Relative energies (eV) of local energy minima on PES
for the adsorbed Sn and Si probe atoms on the stepped Si$(111)-\sqrt{3}\times\sqrt{3}-$Sn
surface. N is the local minimum number according to Fig.~\ref{fig7}. }
\begin{tabular}{ccc}
\multirow{3}{*}{N} & \multicolumn{2}{c}{Relative energy (eV)}\tabularnewline
 & \multirow{2}{*}{Sn } & \multirow{2}{*}{Si }\tabularnewline
 &  & \tabularnewline
1 & 0 & 0\tabularnewline
\multirow{2}{*}{2} & \multirow{2}{*}{-0.33} & \multirow{2}{*}{-0.48}\tabularnewline
 &  & \tabularnewline
3 & -0.58 & -0.55\tabularnewline
4 & -0.28 & -0.61\tabularnewline
5 & -0.19 & -0.51\tabularnewline
6 & -0.07 & -0.23\tabularnewline
7 & -0.37 & -0.52\tabularnewline
8 & -0.49 & -0.68\tabularnewline
9 & -0.39 & -0.47\tabularnewline
10 & -0.18 & -0.23\tabularnewline
\end{tabular}
\end{ruledtabular}

\end{table}

All Si bonds on the Si$(111)-\sqrt{3}\times\sqrt{3}-$Sn surface are
saturated by Sn atoms making this surface inert. Thus, according to
Fig.~\ref{fig7}, each $\sqrt{3}\times\sqrt{3}-$Sn half unit cell
shows only a single shallow local minimum on $\mathrm{T_{4}}$ site
(minimum $\mathrm{N=1}$ ). The energy of this local minimum is taken
as the energy zero in our calculations. The energies calculated for
different energy minima on PES for Sn and Si probe atoms are shown
in Tab.~\ref{tab1}. As expected, the deepest minima are located
near the step edge since this area contains many unsaturated bonds.
According to Fig.~\ref{fig7}, each rest atom is surrounded by four
local minima on PES: $\mathrm{N=2}$, 2, 3, 3 for $\mathrm{R_{2}}$
and $\mathrm{N=7}$, 8, 9, 10 for $\mathrm{R_{1}}$. The region between
$\mathrm{A_{1}}$ and $\mathrm{A_{2}}$ adatoms contains three additional
minima: $\mathrm{N=4}$, 5, 6. The location of the deep local minima
around rest atoms and near adatoms is typical of the surfaces containing
these atoms, and it was also observed in other systems \cite{cha03,zha10}.
The reason for this is that the adsorbed atom in the vicinity of rest
atoms and adatoms may effectively saturate several bonds. According
to Tab.~\ref{tab1}, the two neighboring minima near $\mathrm{R_{2}}$
labeled $\mathrm{N=3}$ on PES are the deepest (global) minima for
the probe Sn atom on the stepped Si$(111)-\sqrt{3}\times\sqrt{3}-$Sn
surface. Therefore, these minima should be the preferential adsorption
sites for Sn atoms at a low sample temperature. 

The calculated PES (Fig.~\ref{fig7}) shows that the energy barrier
between two neighboring $\mathrm{N=3}$ sites is among the lowest
on the surface. As it follows from the symmetry of the computational
cell, the saddle point between these two sites is situated in the
$\left(1\overline{1}0\right)$ plane crossing $\mathrm{A_{1}}$ atom.
The total energy related to this saddle point was calculated by relaxing
the atomic structure while restricting the Sn probe atom movement
to the $\left(1\overline{1}0\right)$ plane. It was found that the
energy barrier between two neighboring $\mathrm{N=3}$ sites is as
low as $0.21\,\mathrm{eV}$. Thus, these two global minima sites are
separated by a low energy barrier and form a double well for the adsorbed
Sn atoms. The calculated energy barrier is close to the $E_{a}$ range
suitable for the observation of the thermally activated dynamical
fluctuations in STM ($0.13-0.15\,\mathrm{eV}$ at 80~K). Therefore,
we suggest that an additional adsorbed Sn atom should flip between
two traps of the double well at 80~K. The irregular bright spots
visible on the lower terrace near the step edge in the $\sigma$ map
(Fig.~\ref{fig6}(a)) originate, most probably, from the random Sn
atoms trapped in the double wells. Note that there is a good correspondence
between the positions of double wells along the step edge (Fig.~\ref{fig7},
between edgemost Sn adatoms on the upper terrace) and the positions
of bright spots in the measured $\sigma$ map (Fig.~\ref{fig6}(a)).
This observation is an additional proof of the developed step model
and proposed interpretation of the experimental data.

\begin{figure}
\includegraphics[width=8.5cm]{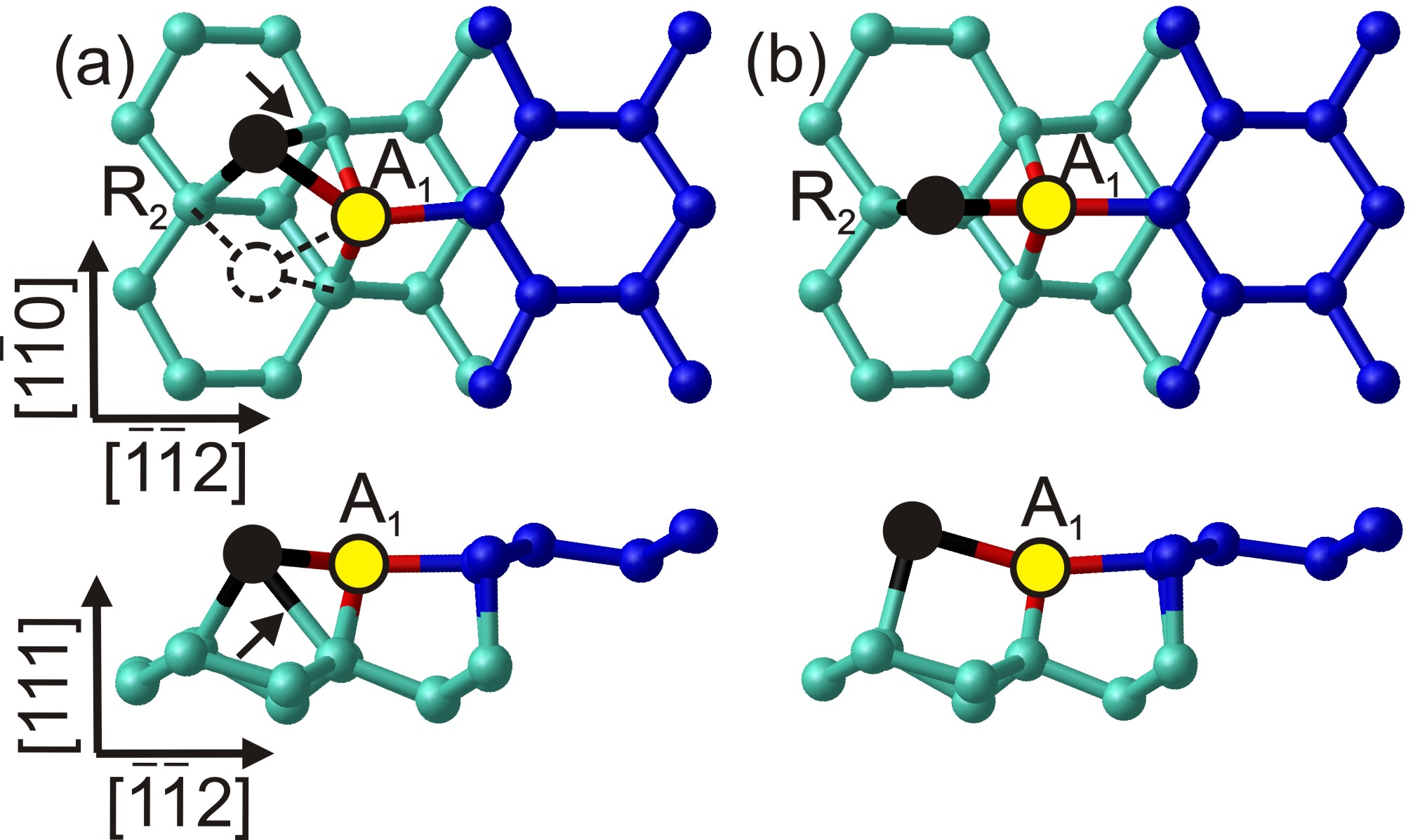}

\caption{\label{fig8}Atomic structures related to the $\mathrm{N=3}$ energy
minima on the calculated PES. (a) Local minima configurations. (b)
Saddle point configuration. Upper figures - top views, lower figures
- corresponding side views. Cyan and blue circles are the Si atoms
of the lower and upper terraces, respectively, the yellow circle is
the $\mathrm{A_{1}}$ Sn adatom at the step edge, black circle is
the probe Sn (Si) adatom. The dashed circle shows the position of
the probe adatom in the second local minima of the double well. The
arrow highlights a weak bond between the probe adatom and the Si substrate. }
\end{figure}

The atomic configuration related to the PES energy minima $\mathrm{N=3}$
in the double well is shown in Fig.~\ref{fig8}(a). According to
this configuration, the adsorbed probe Sn atom is three-fold coordinated.
The first bond is connected to the $\mathrm{R_{2}}$ Si rest atom
on the (111) terrace, the second bond is connected to the $\mathrm{A_{1}}$
Sn adatom of the Sn atomic chain. The third bond (marked with an arrow)
shares its connection to the substrate Si atom with the $\mathrm{A_{1}}$
Sn adatom. The bond length between two Sn atoms is 2.95~Å, being
very close to the calculated bond length in bulk $\alpha$-phase Sn
(2.91~Å). The bond lengths between the $\mathrm{A_{1}}$ Sn adatom
and Si substrate are 2.76~Å, 2.72~Å, 3.02~Å, while, for the probe
Sn atom, they are 2.82~Å and 3.06~Å. The longest and, therefore,
weakened Sn-Si bonds of about 3~Å in each group are connected to
the same five-fold coordinated (overcoordinated) Si substrate atom. When the probe Sn atom
is in its saddle position, as shown in Fig.~\ref{fig8}(b), the weak
bond between this atom and the overcoordinated substrate Si atom is
broken. The height change when moving the probe Sn atom from the position
of energy minima to the saddle point is 0.57~Å.

For the sake of completeness, we also calculated PES for the probe
Si atom on the stepped Si$(111)-\sqrt{3}\times\sqrt{3}-$Sn surface
to check if the observed fluctuations of tunneling current in STM
can be caused by trapped Si atoms. We found that the positions of
energy maxima and minima on PES are very similar to the ones observed
for the probe Sn atom. This is explained by the same number of valence
electrons in Si and Sn atoms. However, the values of calculated local
energy minima, in case of the probe Si atom, are different. It is
shown in Tab.~\ref{tab1} that the global minimum for the probe Si
atom is on the $\mathrm{N=8}$ site, which is on the upper terrace
just behind the $\mathrm{A_{1}}$ adatom (Fig.~\ref{fig7}). This
result suggests that, unlike Sn atoms, Si atoms must accumulate at
the upper terrace edge and be stationary at 80~K. Since this was
not observed in the experimental STM images of step edges (Figs.~\ref{fig1}(a)-\ref{fig1}(c)),
then Si adatoms can be excluded.

\section{CONCLUSIONS}

In summary, the structure of the single step and dynamics of adatoms
on the Si$(111)-\sqrt{3}\times\sqrt{3}-$Sn surface were investigated
by LT-STM at 80~K and \emph{ab initio} calculations. The atomic model
of the $\left[11\overline{2}\right]$ single steps on the Si$(111)-\sqrt{3}\times\sqrt{3}-$Sn
surface was developed. The model contains Sn atomic chains along the
step edges. The adatom dynamics was detected at the lower side of
the step edge in STM at 80~K. The PES calculated using the developed
step model reveals the presence of double wells at the lower side
of the step edge acting as traps for adsorbed Sn atoms. The random
Sn atoms trapped within the double wells and flipping between two
stable states explain the fluctuating current detected in STM at 80~K.
\begin{acknowledgments}
RAZ would like to thank the Novosibirsk State University for providing
the computational resources. Cluster computations and paper writing
were supported by the Russian Science Foundation (project no. 19-72-30023). 
\end{acknowledgments}

\bibliographystyle{apsrev4-1}

\end{document}